\documentclass{article}

\usepackage{framed,multirow}

\usepackage{amssymb}
\usepackage{latexsym}

\usepackage{url}
\usepackage{xcolor}
\definecolor{newcolor}{rgb}{.8,.349,.1}

\usepackage{amsmath,epsfig}

\usepackage{graphicx} 
\usepackage{amsmath} 
\usepackage{amssymb}  
\usepackage{amsthm}
\usepackage{amsfonts}
\usepackage{dsfont}
\usepackage{tikz}
\usepackage{xcolor}
\usepackage[english]{babel}
\usepackage[latin1]{inputenc}
\usepackage{subfigure}
\usepackage{hhline}
\usepackage{url}

\newcommand{\e}{\mathrm{e}}

\newcommand{\R}{\mathbb{R}}

\newtheorem{lemma}{Lemma}
\newtheorem{proposition}{Proposition}
\newtheorem{theorem}{Theorem}
\newtheorem{corollary}{Corollary}

\theoremstyle{remark}

\newtheorem{definition}{Definition}

\usepackage{algorithm,algpseudocode}

\usepackage{algorithmicx}
\algnewcommand{\Inputs}[1]{%
  \State \textbf{Inputs:}
  \Statex \hspace*{\algorithmicindent}\parbox[t]{.8\linewidth}{\raggedright #1}
}
\algnewcommand{\Initialize}[1]{%
  \State \textbf{Initialize:}
  \Statex \hspace*{\algorithmicindent}\parbox[t]{.8\linewidth}{\raggedright #1}
}
\newcommand{\algorithmicbreak}{\textbf{break}}
\newcommand{\Break}{\State \algorithmicbreak}
\title{Analysis of SparseHash: an efficient embedding of set-similarity via sparse projections}
\author{D. Valsesia, S. M. Fosson, C. Ravazzi, T. Bianchi, E. Magli
}

\begin{document}

\maketitle

%
%

%
%

\begin{abstract}
Embeddings provide compact representations of signals in order to perform efficient inference in a wide variety of tasks. In particular, random projections are common tools to construct Euclidean distance-preserving embeddings, while hashing techniques are extensively used to embed set-similarity metrics, such as the Jaccard coefficient. In this letter, we theoretically prove that a class of random projections  based on sparse matrices, called SparseHash, can preserve the Jaccard coefficient between the supports of sparse signals, which can be used to estimate set similarities. Moreover, besides the analysis, we provide an efficient implementation and we test the performance in several numerical experiments, both on synthetic and real datasets.
\end{abstract}



\section{Introduction}\label{sec:introduction}
Retrieving meaningful information from large amounts of data is a complex task, often impossible if those data have to be analyzed in their original domain. For this motivation, compact representations have been studied in different frameworks, ranging from information retrieval (see, e.g., \cite{achlioptas2003database,AndoniLSH2008,Toothpic_TMM}) to signal processing (see, e.g., \cite{Dono06,Cand06b}). 

In information retrieval, hash functions are widely used to map data into compact representations, see, e.g., \cite{wan18} and references therein. Traditional hash functions compress arbitrary data to fixed length representations, preserve exact matches, and minimize collisions between different objects. An important purpose of hashing techniques is the evaluation of the similarity between sets of generic objects. This has many applications in information retrieval. An example is the search of near-duplicate documents: this can be performed by finding the number of bag-of-words shared by different documents, which are said to be near-duplicate if this number overcomes a given threshold.  A popular technique to measure  set similarity is the min-wise hashing (also known as MinHash) proposed by \cite{Broder1997}, and further analysed by \cite{Broder2000,Indyk1999,fast_similiary_sketching,beyond_minhash,exact_weighted_minhash}. MinHash approximately preserves the Jaccard coefficient, which is a popular similarity metric for bag-of-words and similar representations, and is used in a wide range of applications, see, e.g., \cite{Ping_Li_bbit}.

In signal processing, compact signal representations are usually referred to as embeddings, see  \cite{Jacques2013,Boufounos2011Secure,adaptive_embedding}. Formally, an embedding is a transformation that maps a set of signals in a high dimensional space to a set in a lower dimensional space, in such a way that the geometry of the set is approximately preserved. The most famous embedding is probably that proposed by \cite{joh84}, which preserves Euclidean distances using random projections. 

Hash functions and embeddings bear many similarities. For example, a class of efficient indexing techniques known as locality sensitive hashing (LSH) can be constructed using both traditional hash functions and embeddings, based on random projections, as studied by \cite{AndoniLSH2008}. Hence, it is not surprising that results in one field can be exploited to obtain significant advancements in the other field, and vice versa.



A novel 
embedding for the Jaccard coefficient, called SparseHash, was proposed by \cite{sparsehash_icme}. SparseHash builds on the concept of sparsity. A signal $u\in\R^n$  is said to be sparse if it has few non-zero components. The index set of its non-zero components is called {\em support}. SparseHash can efficiently evaluate the similarity between sparse signals, in terms of support overlap. Its rationale is based on recent results by \cite{Bioglio2015,Ravazzi2016,rav18}, which show that the sparsity level of a signal, that is,  the size of its support,  can be efficiently estimated from compressed linear projections obtained through sparse random matrices. 

Sparsity is envisaged also in information retrieval, as several compact representations, e.g., bag-of-words, produce sparse features, see, e.g. \cite{huang2008similarity}. Considering the example of near-duplicate documents, typically each document contains only few bag-of-words with respect to the general vocabulary. Given this observation, we can highlight a duality between sparse signals and sets of generic objects: a set $S\subseteq\Omega=\{1,\ldots,n\}$ can be represented as a signal $u\in\{0,1\}^n$, whose entries are $u_i=1$, if $i\in S$, and $u_i=0$, otherwise, and generally we expect that the number $k$ of ones is much smaller than $n$. Similarity between sets can then be interpreted as the overlap between the supports of sparse signals. In this perspective, SparseHash is a natural alternative to MinHash, which can be considered as benchmark for this kind of problems.

%

This letter extends our preliminary work \cite{sparsehash_icme}. 
We propose two metrics to measure similarity in the embedded domain that depend linearly and nonlinearly, respectively, on the original Jaccard coefficient. Moreover, we present a deeper theoretical analysis of performance in terms of estimation of set similarities. We also introduce a new algorithm that implements SparseHash more efficiently, i.e., with the same asymptotic complexity of the MinHash bottom sketch by \cite{bottomk}. However, compared to the bottom sketch, SparseHash has binary measurements rather than real-valued ones, yielding improved compression efficiency.



The letter is organized as follows. In Section \ref{sec:background}, we introduce MinHash and random projections. In Section\ref{sec:proposed}, we illustrate SparseHash and discuss its implementation. In Section  \ref{sec:theory}, we provide the theoretical analysis. Section  \ref{sec:experimental} is devoted to numerical experiments; conclusions are reported in Section \ref{sec:conclusions}.

\section{Related work}\label{sec:background}
MinHash is a hashing method conceived to preserve  the Jaccard coefficient. Specifically, MinHash generates $m\geq 1$ independent hash functions $h_1,\dots, h_m$ which return an integer value for each element of the original set $S$; then it selects the $m$ minimum values $\min_{u\in S}h_1(u),\dots,\min_{u\in S}h_m(u)$. 
Different variants of MinHash have been devised to obtain more compact storage or lower computational complexity of the hashing operation. In particular, we mention (a) $b$-bit minwise hashing by \cite{Ping_Li_bbit}, which quantizes the hashes over $b$ bits instead of using integers, and (b) bottom-$m$ sketch by \cite{bottomk}, which selects the $m$ smallest values of a single hash function instead of the minima of $m$ independent hash functions; this conceptually substitutes the ``sampling with replacement'' operation performed by MinHash with a ``sampling without replacement''. Clearly, the bottom-$m$ sketch has a lower complexity in computing the hashes. 

We mention that techniques have been proposed that aim at extending MinHash beyond Jaccard similarity, by assigning weights to the elements of the set, see \cite{simminhash,weighted_minhash}. This was motivated by bag-of-words representations where the count associated to each element may carry additional information. In the context of hashing functions, BitShred was also proposed by \cite{bitshred}, however it achieves substantially biased estimates of the Jaccard coefficient when compared to MinHash. Besides, Bloom filters have been proposed by \cite{Bloom}, which are space-efficient representations for set  membership queries. They essentially hash each element of a set into positions of a bit array, and they could also be used to estimate the size of the intersection and union between sets. Bloom filters present some challenges like false positive errors, i.e., wrongly indicating that a non-member element is a member of the set (see \cite{luo18}); a comparison with them is postponed to future work.

Concerning random projections, one of the most famous methods for dimensionality reduction has been introduced by \cite{joh84}, which preserves Euclidean distances. Several extensions have been later proposed, that embed the angle between signals (see \cite{Charikar2002,Jacques2013}) or control the maximum distance that is embedded (see \cite{Boufounos2013}). Finally, we notice that sparse random matrices have received some attention for embedding $\ell_2$ or $\ell_1$ distances in \cite{Li:2006:VSR:1150402.1150436}.    
\vspace{-0.1cm}
\section{SparseHash}\label{sec:proposed}
In this section, we illustrate SparseHash. In particular, we describe how to efficiently implement it, without generating the projection matrix, and we propose a novel faster approximated implementation. In the following, we denote the support of $u\in\R^{n}$ as 
$\mathrm{supp}(u)=\{i\in\{1,\ldots,n\}: u_i\neq0\}$; $k$ is the sparsity level, that is, the cardinality of $\mathrm{supp}(u)$, and $\Sigma_k:=\{u\in\R^n:|\mathrm{supp}(u)|\leq k\}$.


We call random projection an algorithm that projects a vector $u\in\R^n$  onto a lower-dimensional subspace $\R^m$ by multiplying it by a random matrix $A\in\R^{m\times n}$, $m<n$, see \cite{Dono06}. The obtained vector $y=Au$ is referred to as measurement, and $\R^m$ is known as reduced space. The intuitive idea is that a properly designed random mapping projects data points onto a randomly selected subspace approximately preserving distances. Generally, dense random matrices are considered in the literature. 

SparseHash consists of computing binary-quantized sparse random projections $y=|\text{sign}(Au)|$, where $A\in\R^{m\times n}$ is a $\gamma$-sparsified random matrix, defined as follows: with probability $1-\gamma$, $A_{ij}=0$; with probability $\gamma$, $A_{i,j}$ is generated according to an arbitrary continuous distribution with zero mean and finite variance.
$\gamma$-sparsified random matrices are efficient for sparsity estimation, as proven by \cite{Ravazzi2016}; their use is the core of SparseHash, and is the basis to provide a rigorous  analysis of its efficiency.
SparseHash can be used to evaluate the similarity of sets of generic objects because, as already mentioned, a set $S\subseteq\Omega=\{1,\ldots,n\}$ that contains $k\ll n$ objects can be represented by a sparse signal $u\in\{0,1\}^n$, and set similarity can be interpreted as support overlap. 

\subsection{Computation via hashing}\label{sec:cvh}
From a practical viewpoint, computing the measurements in SparseHash does not require to explicitly generate $A$ and perform the matrix-vector product. An efficient implementation is possible by using hash functions, see \cite{universal_hash}, which map an index in $S$ to a uniformly distributed value over the output range of the hash function (e.g., $b=64$ bits integers). Let $f_i$ be a hash function that maps its input to an integer in the range $[0,2^b-1]$ and define a threshold $\tau$, where $\gamma$ is as defined in the previous section. A measurement $y_i$ is zero if and only if the hash function returns a value below $\tau$ for all the indexes in the support. By defining $\tau=\gamma (2^b-1)$, measurements have the same probability to be zero as in the formulation $y=|sign(Au)|$ illustrated in the previous paragraph.

Multiple hash functions are used to generate $m$ measurements. This typically involves randomizing a seed of the hash function. Algorithm \ref{alg:sparsehash} summarizes all the steps required to generate the SparseHash measurements $y$. 
\begin{algorithm}[h]
\begin{algorithmic}
{\small{
\Inputs{$\gamma$, $S=\mathrm{supp}(u)=\lbrace s_j \rbrace_{j=1}^k$}
\Initialize{$\tau \gets \gamma \left( 2^b-1 \right)$; $~~~y_i \gets 0$, $i=1,\ldots,m$}
\For{$i = 1, \ldots, m$}
    \For{$j = 1, \ldots, k$}
    \State $h_{ij} \gets f_i(s_j)$
    \If{$h_{ij} < \tau$}
        \State $y_i \gets 1$
        \Break
    \EndIf
    \EndFor
\EndFor
}}
\end{algorithmic}
\caption{Computing SparseHash}
\label{alg:sparsehash}
\end{algorithm}

Algorithm \ref{alg:sparsehash} is equivalent to computing $y=\vert \mathrm{sign}(Au) \vert$ in the sense that it approximately yields the same probability to get a nonzero measurement $y_i$ as function of the size of the support. The equivalence would be exact if the hash function could generate output values that are truly uniformly distributed and whose range is large enough that that the quantization of probabilities is negligible. However, popular functions, e.g., \cite{murmurhash3}, are designed to be as uniform as possible and work using 32 bits or more as output range, which is large enough to be a good approximation. Finally, the procedure is repeatable, i.e., an hash function returns the same output to the same input.

%
\begin{algorithm}[p]
\begin{algorithmic}
{\small{
\Inputs{$\gamma$, $S=\mathrm{supp}(u)=\lbrace s_j \rbrace_{j=1}^k$}
\Initialize{$\tau \gets \gamma \left( 2^b-1 \right)$; $~~y_i \gets 0$, $i=1,\ldots,m$}
\For{$j = 1, \ldots, \vert S \vert$}
    \State $h_{j} \gets f(s_j)$ \Comment{Compute hashes}
\EndFor
\For{$i = 1, \ldots, m$}
    \State Generate random bot[i]\Comment{Generate bottom values} 
\EndFor
\State Sort bot
\State head $\gets$ buildTree(bot)
\For{$j = 1, \ldots, k$}
\State ptr$\gets$head
\While{ptr $\neq$ NULL}
    \If{$h_j < $ ptr.bottomValue}
        \State ptr $\gets$ ptr$.$left
    \Else
        \If{$h_j < $ ptr.topValue}
            \State $i \gets $ptr.measIndex; $~~~y_i \gets 1$
            \If{ptr$.$measIndex $\neq m-1$} \Comment{Check right}
                \State $p \gets$ ptr.measIndex $+1$
                \While{$p < m$ and $h_j \geq $bot[$p$]}
                    \State $y_p \gets 1$; $~~p \gets p+1$
                \EndWhile
            \EndIf
            \If{ptr.measIndex $\neq 0$} \Comment{Check left}
                \State $p \gets$ ptr.measIndex $-1$
                \While{$p \geq 0$ and $h_j < $bot[$p$]$+\tau$}
                    \State $y_p \gets 1$; $~~p \gets p-1$
                \EndWhile
            \EndIf
            \Break \Comment{No more measurements can collide}
        \Else
            \State ptr $\gets$ ptr.right
        \EndIf
    \EndIf
\EndWhile
\EndFor
}}
\end{algorithmic}
\caption{Computing Fast SparseHash}
\label{alg:fastsparsehash}
\end{algorithm}

\begin{figure}[h]
    \centering
    \includegraphics[width=0.62\columnwidth]{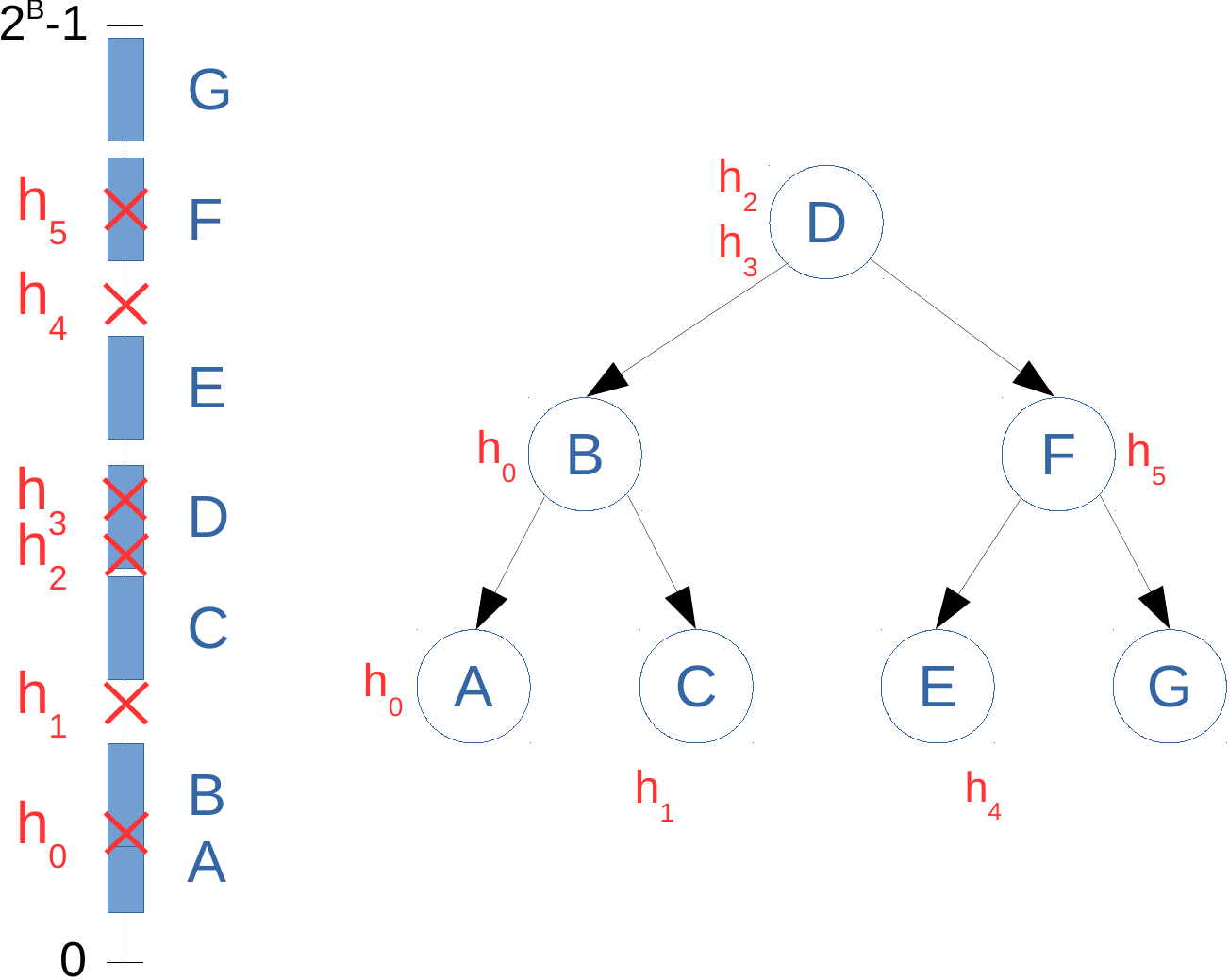}
    \caption{Example of Fast SparseHash: a binary tree is generated from $m=7$ random windows, denoted as A to G. The membership of the $k=6$ hashes in the windows are determined by traversing the tree. The corresponding  SparseHash measurements are 1101010. The tree structure must have a \textsc{right} and a \textsc{left} pointers, and an integer \textsc{measIndex} in $[0,m-1]$ storing the index of the measurement.}
    \label{fig:tree}
\end{figure}
\subsection{Fast SparseHash}\label{sec:fast_sparsehash}
The technique to implement SparseHash just described requires to compute $\mathcal{O}(km)$ hashes (by hash, we mean the output of the hash function applied to a single entry of the support), and perform $\mathcal{O}(km)$ comparisons with the threshold $\tau$. The complexity is thus equivalent to that of MinHash, which  computes $\mathcal{O}(km)$ hashes and performs $\mathcal{O}(km)$ operations to find the $m$ minima. However, there are variants of MinHash, in particular the bottom-$m$ variant by \cite{bottomk}, which reduce the complexity. SparseHash can be modified so as to require only $\mathcal{O}(k)$ hashes and $\mathcal{O}(k \log m)$ comparisons. We now illustrate this variant, that we call Fast SparseHash. 
The main idea behind Algorithm \ref{alg:fastsparsehash} is to compute only one hash per support entry and to check if it falls inside one of $m$ randomly drawn windows of width $\tau$ instead of falling below a fixed threshold. A single hash is computed for each of the $k$ elements in the support and $m$ random windows of width $\tau=\gamma(2^b-1)$ are drawn over the output range of the hash function. Notice that the width $\tau$ of the window is exactly the same as in Algorithm \ref{alg:sparsehash}. A naive solution to compute the measurements would consist in setting a measurement to 1 if at least one of the hashes falls inside the window drawn for that measurement. However, this solution would still require $\mathcal{O}(km)$ comparisons. A better solution is to use a binary search tree to store the windows by means of their sorted bottom values so that the value of the measurement can be determined by traversing the tree, yielding a logarithmic complexity in the number of measurements. More precisely, the tree stores the measurement number $i \in \left[ 0,m-1 \right]$, the value of the bottom of the corresponding window as well as pointers to the two children. The tree is created in the following way. The bottom values are sorted in increasing order and their median value is inserted as root of the tree. Then the tree is recursively created following the rule that the left (respectively, right) subtree includes the windows with bottom values smaller (respectively, larger) than the parent. To determine whether the measurements are zero or nonzero, each hash of the set traverses the tree. The hash is first compared to the value of the window bottom stored in the root node to determine if it falls inside that window. If it does, the measurement corresponding to the index stored in that node is set to 1 and the next hash is examined. Otherwise, the right or left children is examined depending if the hash is below or above the bottom of the current window. This is repeated until a leaf is reached, or a measurement is set to 1. Partially overlapping windows are handled with a local search: if a hash is determined to fall inside a window, the windows whose bottoms are immediately smaller or larger are checked to determine if the hash also falls inside them. This is repeated until no neighboring windows report the hash falling inside. Algorithm \ref{alg:fastsparsehash} shows the whole procedure to compute measurements with Fast SparseHash, while Figure \ref{fig:tree} shows the process in a pictorial fashion. Table \ref{table:speed} reports an experimental comparison of the runtime for SparseHash and Fast SparseHash for various values of $k$ and $m$: Fast SparseHash is significantly faster and has a sublinear increase in runtime for increasing $m$ while SparseHash has a linear increase.
\begin{table}[t]
\caption{Fast SparseHash v. SparseHash - Runtime (sec.)}
\footnotesize
\centering
\begin{tabular}{cc|ccccc}
& & \multicolumn{3}{c}{$m$}\\
& &  $10^4$          & $10^5$           & $10^6$           \\
 \hline
&$10^4$                   & \textbf{0.007} / 1.337   & \textbf{0.040} / 12.36   & \textbf{0.194} / 118.8  \\
$k$&$10^5$                   & \textbf{0.023} / 11.26  & \textbf{0.061} / 115.5  & \textbf{0.280} / 1119 \\
&$10^6$                   & \textbf{0.136} / 126.8 & \textbf{0.205} / 1163 & \textbf{0.476} / 11129        
\end{tabular}
\label{table:speed}
\vspace{-0.3cm}
\end{table}
\vspace{-0.2cm}

\section{Analytical results}\label{sec:theory}
In the following, we present two metrics to compute set similarities from measurements in the reduced domain. 
Given $u,v\in\mathbb{R}^n$ and their measurements $y,z \in \lbrace 0,1 \rbrace^m$ obtained via SparseHash, we are interested in defining a similarity metric between $y$ and $z$ that approximately embeds the Jaccard coefficient $J(S_u,S_v)=|S_u\cap S_v|/|S_u\cup S_v|$ of $S_u=\mathrm{supp}(u)$ and $S_v=\mathrm{supp}(v)$. To simplify the notation, let $J_{u,v}:=J(S_u,S_v)$. 
\subsection{Jaccard coefficient}\label{sec:simsh}
Let $y,z \in \lbrace 0,1 \rbrace^m$. We define
$$\text{sim}_{\cup}(y,z):=\frac{1}{m}\sum_{i=1}^m\mathds{1}({\{y_i=0,z_i=0\}}),$$
$$\text{sim}_{\cap}(y,z):=\frac{\sum_{i=1}^m\mathds{1}({\{y_i=0\})\sum_{j=1}^m\mathds{1}(\{z_j=0\}})}{m\sum_{i=1}^m\mathds{1}({\{y_i=0,z_i=0\}})}$$
where $\mathds{1}({\{A\}})$ is the indicator function which returns 1 when $A$ is true, while $\mathds{1}({\{A,B\}})$ returns 1 when both A and B are true. Then, comparison in the reduced space can be done with the following similarity index:
\begin{equation}
\label{eq:similarity}
\text{sim}_{\sf{sh}}(y,z):=\frac{\log(\text{sim}_{\cap}(y,z))}{\log(\text{sim}_{\cup}(y,z))}.
\end{equation}
We notice that $\gamma$ must be designed so that this formula has small probability to be undefined. In the following theorem, we state that $\text{sim}_{\sf{sh}}$ is a random variable that concentrates around the Jaccard coefficient between the supports of the original signals. We emphasize that Theorem 1 is a more refined version of Proposition 1 in \cite{sparsehash_icme}. More precisely, a deeper theoretical analysis leads to a new estimation of the performance that is more explicit in terms of the main parameters $m,\gamma,k_{\min},k_{\max}$ and $N$.
\begin{theorem}\label{thm: concentration_Jaccard}
Let $\mathcal{X}_{N}=\{x_i\in \R^n:|\mathrm{supp}(x_i)|\in[k_{\min},k_{\max}]\}_{i=1}^N$ be a set of $N$ sparse vectors. For any $\epsilon>0, \beta>2$ and any integer $n$, let $m$ be a positive integer such that
$m> 32\frac{\log 4+\beta\log N}{\gamma^2k^2_{\min}\e^{-\gamma k_{\max}}\epsilon^2}$.
Then,
\begin{align}
\label{eq:expected_jaccard}
\mathbb{P}\left[\bigcup_{(u,v)\in\mathcal{X}_N}\left\{|\mathrm{sim}_{\sf{sh}}(Au,Av)- J_{u,v}|>\epsilon\right\}\right]\leq N^{-\beta+2}.
\end{align}
\end{theorem}
\textit{Sketch of the proof} {\color{black} For brevity, we report only the key steps of the proof; details are provided in the supplementary material}. Let $u\in\Sigma_{k_1}$ and $v\in\Sigma_{k_2}$. Let us define: $\zeta:=(1-\gamma)^{\frac{k_1+k_2}{1+J_{u,v}}}$.
Exploiting the Hoeffding's inequality (see \cite{citeulike:3392582}), we can prove the following inequalities:
\begin{equation}\label{e1}
\begin{split}
&\mathbb{P}\left(\left|\mathrm{sim}_{\cup}(Au,Av)-\zeta]\right|>\epsilon\right)\leq 2\e^{-2\epsilon^2m}\\
%
&\mathbb{P}\left(\left|\mathrm{sim}_{\cap}(Au,Av)-\zeta^{J_{u,v}}]\right|>\epsilon\right)\leq 6\e^{-m\min\left\{(1-\gamma)^{8k_{\max}}\frac{\epsilon^2}{8},\frac{(1-\gamma)^{4k}}{2}\right\}}.
\end{split}
\end{equation}
By using the fact that for any positive random variable $X$  such that $\mathbb{P}\left(|X-\mu_X|>\epsilon\right)\leq p_X(\epsilon)$ with $\mu_X>0$, it holds that 
$\mathbb{P}\left(\left|\log(X)-\log(\mu_X)\right|>\epsilon\right)\leq p_X\left(\epsilon\mu_X\right)$, we can deduce the following inequalities from \eqref{e1}:
\begin{equation}
    \begin{split}\label{e4}
&\mathbb{P}\left(\left|\log(\mathrm{sim}_{\cap}(Au,Av))-J_{u,v}\log\zeta\right|>\epsilon\right)\leq 6\e^{-m(1-\gamma)^{12k_{\max}}\frac{\epsilon^2}{8}};\\
&\mathbb{P}\left(\left|\log(\mathrm{sim}_{\cup}(Au,Av))-\log\zeta\right|>\epsilon\right)\leq  2\e^{-2m\epsilon^2(1-\gamma)^{4k_{\max}}}.
\end{split}
\end{equation}
Moreover, we can prove that, given $X_u=\frac{\sum_{i=1}^m\mathds{1}(\{(Au)_i=0\})}{m}$,  \begin{equation}\label{e5}
\mathbb{P}\left(\left|X_u X_z-(1-\gamma)^{k_1+k_2}\right|>\epsilon\right)\leq 4\e^{-2m\min\left\{\frac{\epsilon^2}{9},\frac{(1-\gamma)^{2k_{\max}}}{4}\right\}}.
\end{equation}
Finally, by merging \eqref{e4} and \eqref{e5}, under the given assumption on $m$, 
the thesis is easily derived.\qed

For large $n$, $k$, and $m$, $\gamma k$ represents
the average number of nonzeros in each row of $A$ that align with the support of cardinality $k$. This observation reveals three regimes, corresponding to the scaling of $\gamma$ and $k$: if $\gamma k_{\min}=\Theta(1)$, then $m=O(\log N)$ is sufficient to get the bound in Theorem \ref{thm: concentration_Jaccard}. In sharp contrast, if $\gamma k_{\max}\rightarrow\infty$ or $\gamma k_{\min}\rightarrow 0$, then $m$ must increase to guarantee the concentration with high probability.

\subsection{Hamming distance and LSH}\label{sec:dham}
As for MinHash, signals can be compared in the reduced space using the Hamming distance $d_{\sf{H}}$ between two hash codes.
From the law of large numbers we have the following properties. Given $u\in\Sigma_{k_1}$ and $v\in\Sigma_{k_2}$
{\small{\begin{equation}\label{eq:ham_formula}
E_{\text{sh}}:=\frac{\mathbb{E}[d_{\sf{H}}(Au,Av)]}{m}=(1-\gamma)^{k_1}+(1-\gamma)^{k_2}-2(1-\gamma)^{\frac{k_1+k_2}{1+J_{u,v}}}
\end{equation}}}
and 
$
\widehat{V}:=\mathsf{Var}[d_{\sf{H}}(Au,Av)/m]= (1-E_{\text{sh}})E_{\text{sh}}/{m}.
$
 For signals with similar sparsity degree $k_1\approx k_2=k$, by setting $(1-\gamma)^k=1/2$ in order to maximize the entropy of the binary measurements, we obtain: 
$E_{\text{sh}}\approx1-2^{\frac{J-1}{1+J}}.$
The characterization of the relationship between the Hamming distance of the hashes and the original Jaccard coefficient derived in \eqref{eq:ham_formula} is important in the context of LSH. In a nutshell, LSH allows us to  approximate nearest neighbor database searches with sublinear complexity (see \cite{AndoniLSH2008}),  without scanning all the entries in the database. Let $\mathcal{X}\subset\R^n$ be a set of points with distance measure $d_{\mathcal{X}}$, and consider the $(R,c)$-NN problem where one is concerned with retrieving all the neighbors of the query point within a distance $R$, while discarding the points at distances greater than $cR$. An LSH family is defined as follows.
\begin{definition}
Let $p_1> p_2$ and $r_1<r_2$. A family $\mathcal{H}: \{h: \mathcal{X}\rightarrow\mathcal{U}\}$ is called $(r_1,r_2,p_1,p_2)$-sensitive for $d_{\mathcal{X}}$ if for any $x,\xi\in\mathcal{X}$ the following fact hold: if $d_{\mathcal{X}}( x,\xi) \leq r_1$ then $\mathbb{P}\left(h(\xi)=h(x)\right)\geq p_1$;  if $d_{\mathcal{X}}( x,\xi) \geq r_2$ then $\mathbb{P}\left(h(\xi)=h(x)\right)\leq p_2$.
\end{definition}
In the $(c,R)$-NN problem, we set $r_1=R$ and $r_2=cR$ and we define a new family of functions
$
\mathcal{F}=\{f: \mathcal{X}\rightarrow \mathcal{U}^m\}
$
such that
$
f(x)=(h_1(x),\ldots,h_{m}(x))^{\top}
$, 
where $h_i\in\mathcal{H}$ are chosen independently uniformly at random from $\mathcal{H}$. We notice that $f\in\mathcal{F}$ is $(r_1,r_2,p_1^m,p_2^m)$-sensitive for $d_{\mathcal{X}}$.
Fixed $L>0$, we define a new family $\mathcal G$ of hash functions $g$ constructed from $L$ random functions $f_1,\dots, f_L$ from $\mathcal F$. We say that $g(\xi)=g(x)$ if $f_i(\xi) = f_i(x)$ for at least one $i \in \{1,\ldots,L\}$. Since the members of $\mathcal F$ are independently chosen for any $g \in \mathcal G$, $\mathcal G$ is a $(r_1, r_2, 1- (1 - p_1^m)^L, 1 - (1 - p_2^m)^L)$-sensitive family. 

During preprocessing, $L$ hash tables, each corresponding to a different hash function $g_i$, are constructed, by storing each $x\in\mathcal{X}$ in the bucket $g_i(x)$. Given a query item $\xi$, we retrieve first $3L$ data points that are hashed to the same bucket $g_i(\xi)$ with $i=1,\ldots, L$ and if there is a point $x^{\star}$ within distance $cR$ from $\xi$ we return 'yes' and $x^{\star}$, else we return 'no'.
If $   m={\log N}/{\log (1/p_2)}, L=N^{\rho}, \rho=\frac{\log(1/p_1)}{\log(1/p_2)}$
then the algorithm is successful with constant probability and the algorithm has the following properties: (a) preprocessing time is $O(N^{1+\rho} m T)$, where $T$ is the time to evaluate a function $h \in \mathcal H$ on an item; (b) storage is of order $O(NL+Nm)=O(N^{1+\rho}+Nm)$; (c) query time is $O(L(m T+nNp_2^m))=O(N^{\rho}(mT+n))$.
Notice that savings in terms of storage can be achieved when 1-bit  measurements are used in place of real-valued measurements.

We now study the performance of LSH in terms of storage and time requirements to respond to a query, using MinHash and SparseHash as embeddings. Such performance metrics are entirely governed by embedding, more precisely, by the function that maps the Jaccard coefficient to the probability of having two equal bits. This function is linear in MinHash and nonlinear in SparseHash. In order to make a fair comparison, we use 1-bit  MinHash, so that the storage complexity is equalized between the two approaches. Notice that $m$ is chosen so as to minimize $L$. 
If a pair of signals in $\mathcal{X}$ has Jaccard coefficient $J$, then the probability that their hashes computed with SparseHash become a candidate pair is given by:
$P_{\text{sh}}=1-(1-p_{\text{sh}}^{m_{\text{sh}}})^{L_{\text{sh}}}$
with $p_{\text{sh}}=2^{\frac{J-1}{1+J}}$, while 
for 1-bit MinHash,  
$P_{\text{mh}}=1-(1-p_{\text{mh}}^{m_{\text{mh}}})^{L_{\text{mh}}}$ 
with $p_{\text{mh}}=(J+1)/2$. The following proposition states that, with equal $m$, SparseHash requires less tables, which enables  shorter query times and lower storage requirements.
%
%
\begin{proposition}\label{prop:lsh_sh_vs_minhash}
If $P_{\text{mh}}=P_{\text{sh}},\  m_{\text{mh}}=m_{\text{sh}}=m
$, then
${L_{\text{sh}}}\leq {L_{\text{mh}}}.$
\end{proposition}
\begin{proof}
We have $$\frac{L_{\text{sh}}}{L_{\text{mh}}}
  =\frac{\log\left(1-\left((J+1)/2\right)^{m}\right)}{\log\left(1-2^{m\frac{J-1}{1+J}}\right)}.$$ We then obtain $L_{sh}\leq L_{mh}$ if $2^{\frac{J-1}{1+J}}\geq (J+1)/2$, which holds for any $J\in [0,1]$. More details are provided in the supplementary material.
\end{proof}
%
%

\section{Numerical experiments}\label{sec:experimental}
In this section, we propose numerical experiments that validate the theoretical results and show the effective performance of SparseHash, both on synthetic and real datasets. The code for these experiments is available in \cite{sparsehash_code}.
\subsection{Numerical validation}\label{sub:nv}
In \cite{sparsehash_icme}, a numerical validation on the concentration of SparseHash around the true Jaccard coefficient was proposed. We retrieve the same experiment to validate the concentration on the Hamming distance.
\begin{figure}[ht]
    \centering
    \includegraphics[width=0.58\columnwidth]{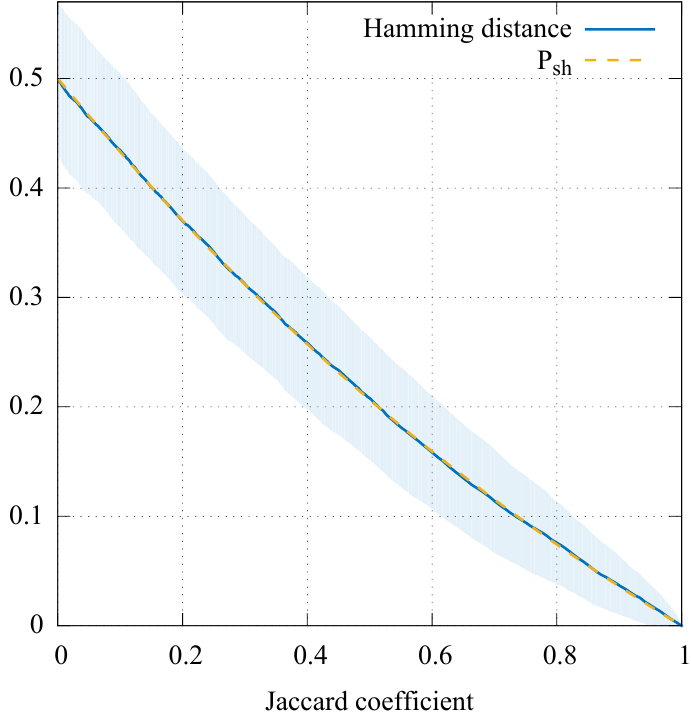}
    \caption{Numerical validation: Hamming distance (mean and variance).} 
    \label{fig:hamming}
\end{figure}
We randomly generate a large number of signals with different amount of support overlap and compute their random projections via $\gamma$-sparsified matrices $A\in\R^{m,n}$. We set $n=1000$, $m=50$, and sparsity level $k=230$. Mean and variance are evaluated over 500  runs.
$\gamma$ is set as the value that maximizes the entropy of the binary measurements, i.e. generates zero or nonzero measurements with equal probability. Since $\mathbb{P}(f_i(u)=0)=(1-\gamma)^k$, we set:
$\gamma = 1-2^{-\frac{1}{k}}\approx 3\cdot10^{-3}.$
In Figure \ref{fig:hamming}, we depict the Hamming distance. The dashed cyan line is the theoretical mean $P_{sh}$ computed from the true Jaccard coefficient defined in \eqref{eq:ham_formula}. As expected, the experimental mean overlies the theoretical mean.
\subsection{Similar text documents}
\begin{figure*}
    \centering
    \includegraphics[width=0.681\columnwidth]{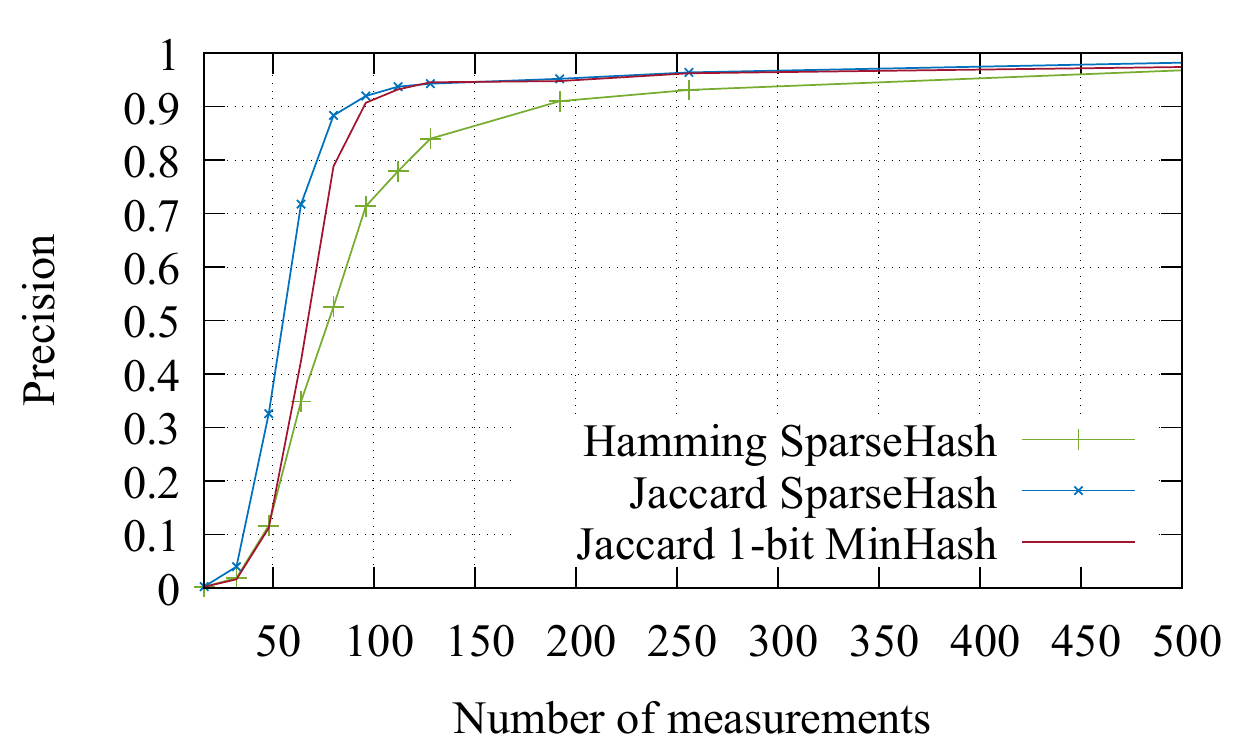}
    \includegraphics[width=0.681\columnwidth]{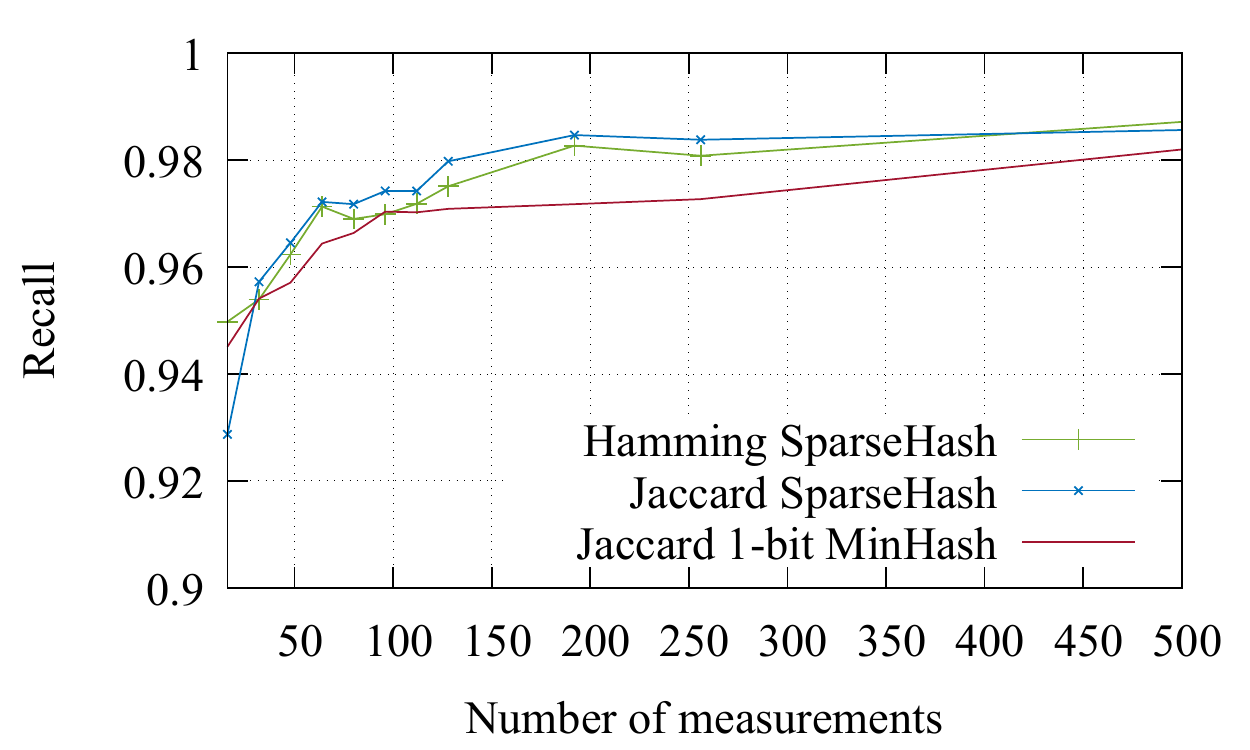}
     \includegraphics[width=0.681\columnwidth]{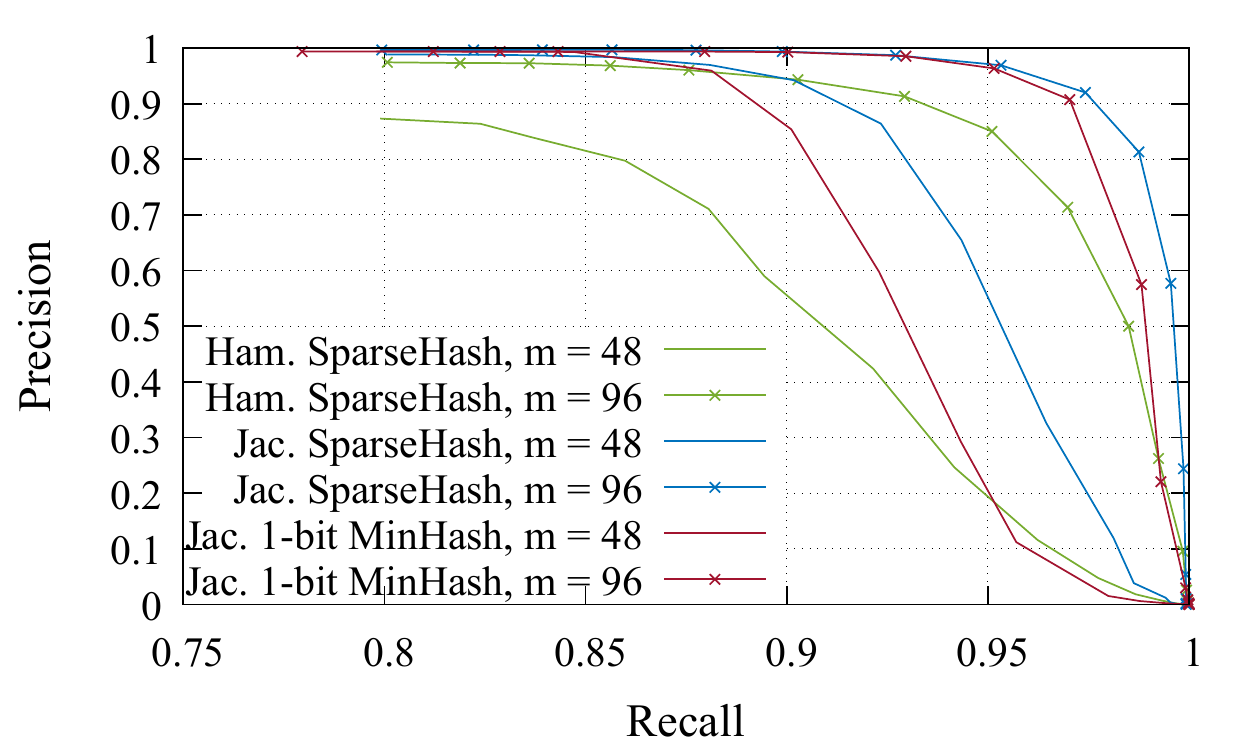}
    \caption{Experiment on similar text documents: precision and recall, threshold 0.5.}
    \label{fig:precrec05}
\end{figure*}
\begin{figure*}
    \centering
    \includegraphics[width=0.681\columnwidth]{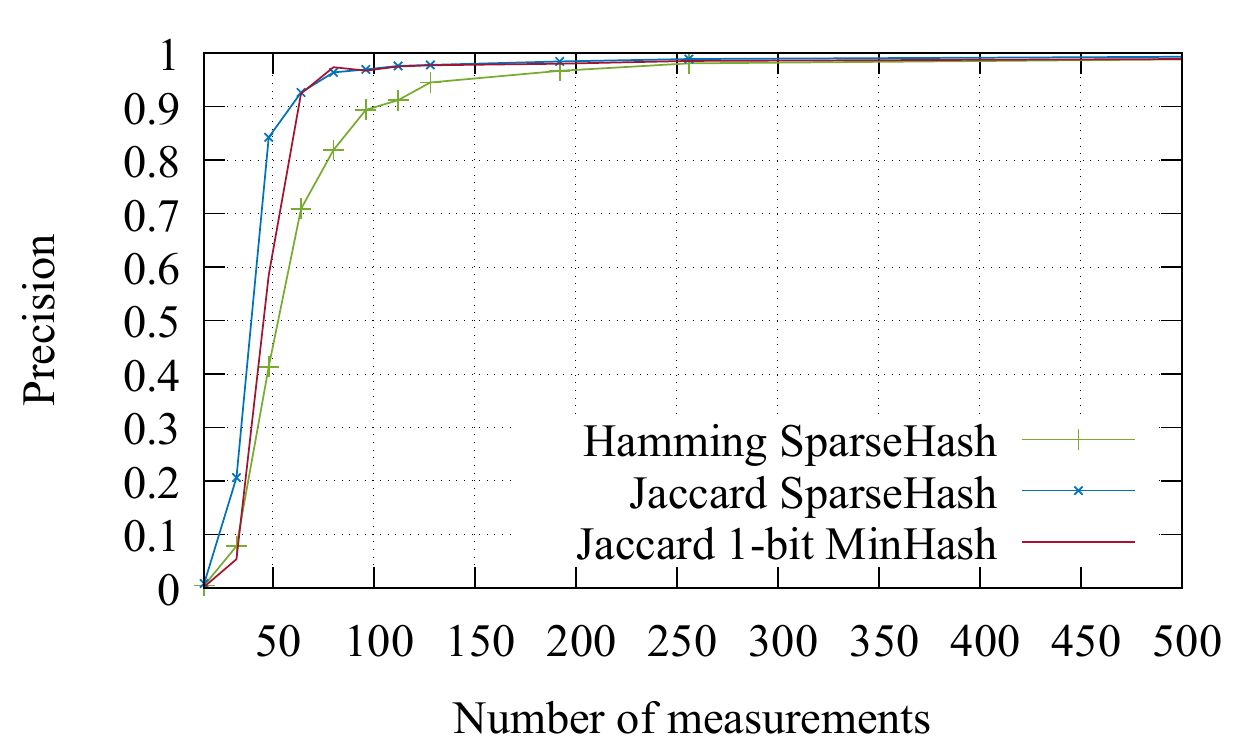}
    \includegraphics[width=0.681\columnwidth]{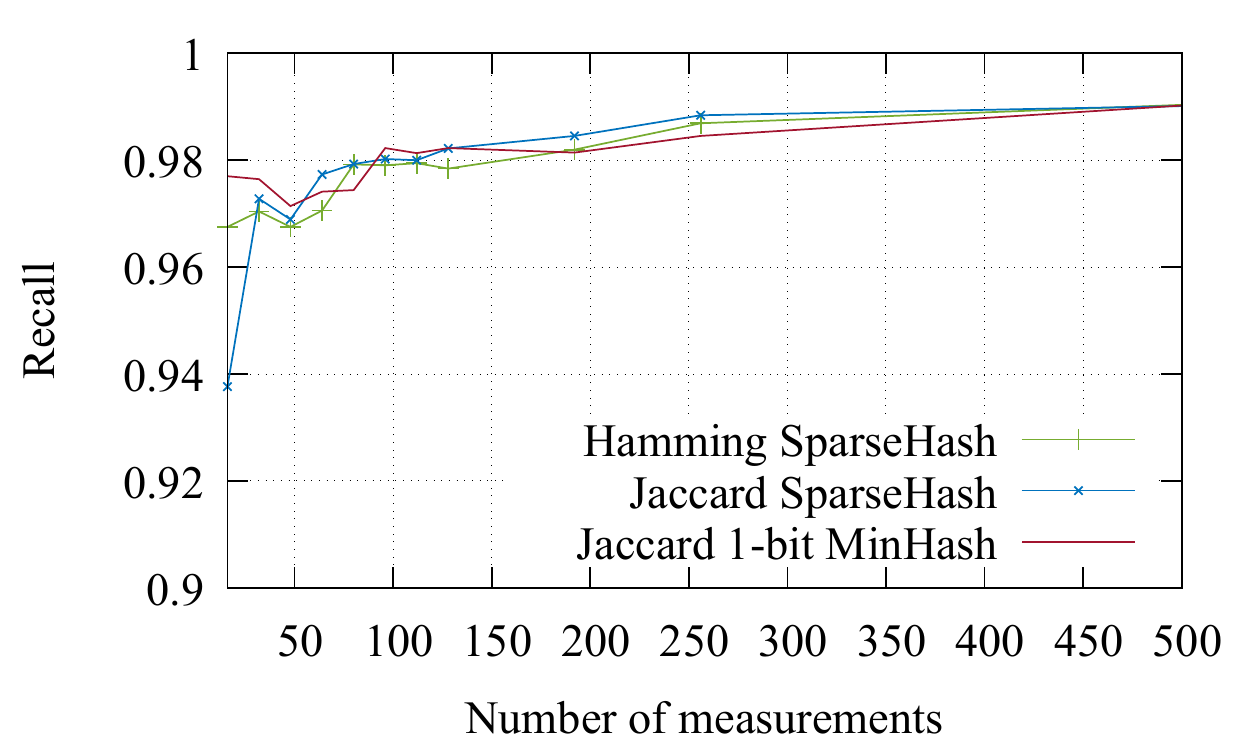}
    \includegraphics[width=0.681\columnwidth]{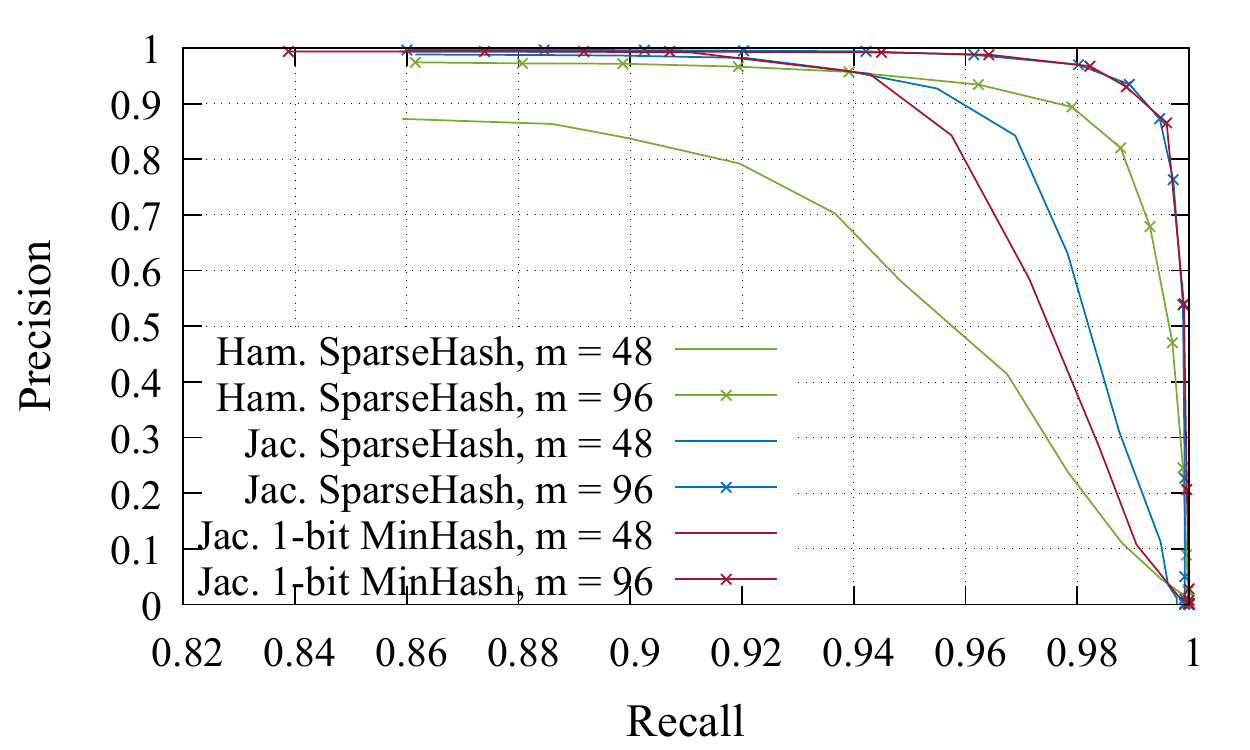}
    \caption{Experiment on similar text documents: precision and recall, threshold 0.6.}
    \label{fig:precrec06}
\end{figure*}
As discussed in \cite{sparsehash_icme}, the problem of finding near-duplicate or similar documents in an archive of text data is outstanding (see \cite{Broder1997,Broder1997clustering,Henzinger2006}). Documents can be represented with bag-of-words models. Given a vocabulary of $n$ words, we can associate a document with a $u\in\R^n$, where $u_i$ counts the occurrences of the $i$th word of the vocabulary in the document. Bag-of-words models typically yield  sparse signals, as the number of different words appearing in a single document is usually much smaller than the size of the vocabulary. We now retrieve the experiment proposed in \cite{Ping_Li_bbit}, where the effects of quantization are evaluated on MinHash. We use the UCI dataset of New York Times articles (see \cite{NYTimes}), composed of about 300000 news articles, with a bag-of-words representation given for each article. The vocabulary contains $n=102660$ words.
The mean he number of different words used in each article is $k=232$; we use this value to assess $\gamma$ as set in Section \ref{sub:nv}.

We compare the performance of SparseHash (with both Jaccard and Hamming metrics) and 1-bit MinHash, in terms of precision and recall. Specifically, we define as similar the documents with Jaccard coefficient larger than a certain threshold, and we try to detect them. In figures \ref{fig:precrec05} and \ref{fig:precrec06}, we set the threshold to 0.5 and 0.6, respectively, and we show precision and recall as functions of the number of measurements $m$, and precision as a function of recall. We see that the precision of SparseHash with Jaccard metric outperforms 1-bit MinHash, in particular when $m<100$. For larger $m$, both methods are efficient, with precision close to 1. The recall is close to 1 for both methods. The gain obtained by SparseHash is well visualized also in the precision-recall curves, depicted for $m=48$ and $m=96$. 
\subsection{Metagenome clustering}
Metagenome clustering is concerned with detecting communities of microorganisms starting from genomic sequences. A genomic sequence can be seen as a long string of A, C, T or G characters representing the four types of nucleotides. In the following, we deal with assembled sequences, which are reconstructed from overlapping partial reads produced by the sequencing instruments. Metagenome clustering is formulated in terms of pairwise distances between sequences. The distance metric of interest is 1-ANI: ANI is the average nucleotide identity, i.e., the percentage of unchanged nucleotides in the two sequences. Since the genomic sequences can be extremely long, dimensionality reduction methods are essential to efficiently compute distances. In \cite{Ondov2016}, the so-called MASH algorithm uses MinHash to this purpose. The authors split a sequence into substrings, called $\kappa$-mers, using a sliding window approach. In their experiments, $\kappa=21$. Each genomic sequence is then represented by the set of its $\kappa$-mers and the Jaccard coefficient between such sets correlates with the expected ANI. By using MinHash to compress the set of $\kappa$-mers, the authors achieve a significant dimensionality reduction.

\begin{figure}
    \centering
    \includegraphics[width=0.53\columnwidth]{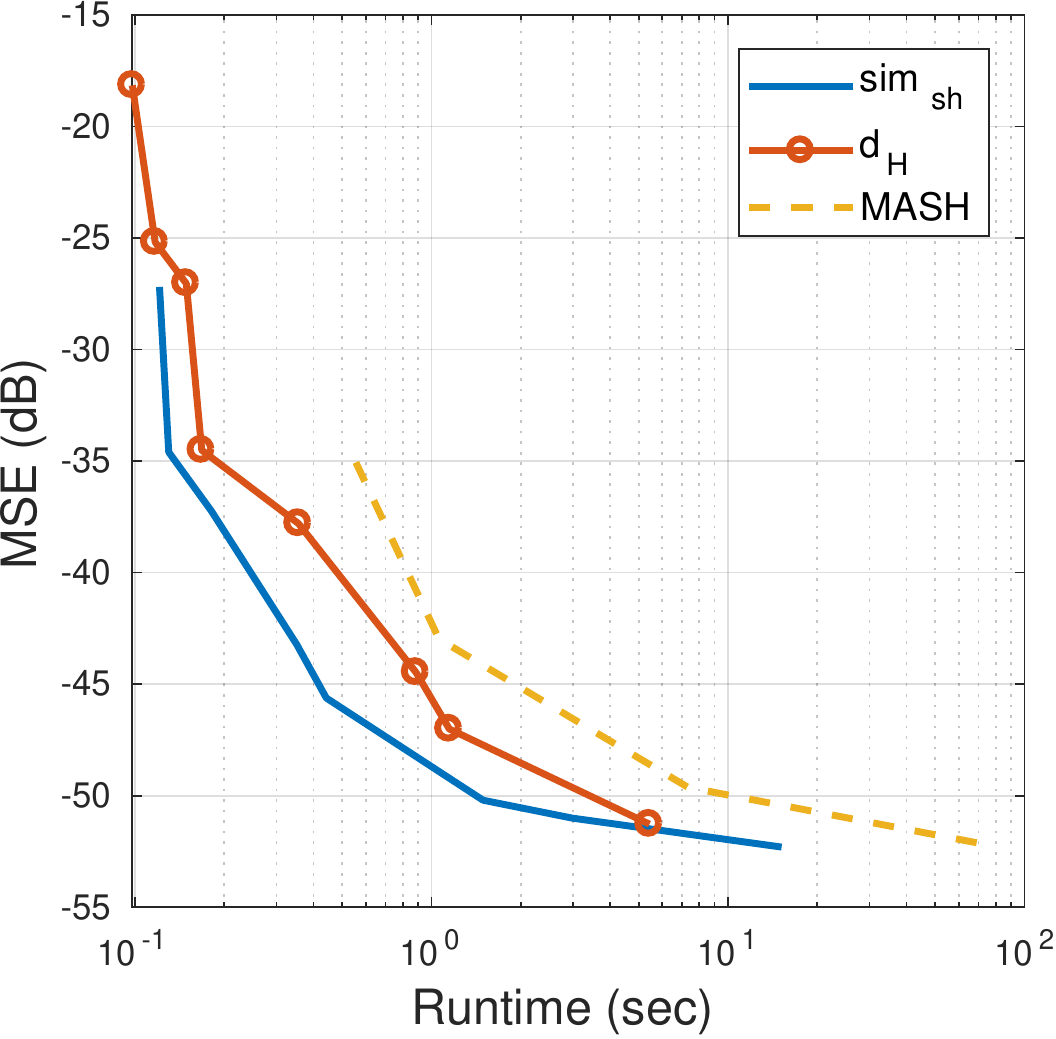}
    \caption{Metagenome clustering: MSE on pairwise Jaccard matrix with computation time.}
    \label{fig:metagenome}
\end{figure}

In this section, we test SparseHash for the same purpose, and compare it to MinHash. The performance is evaluated on the capacity of preservation of the Jaccard coefficient. The dataset from \cite{human2012structure} is used, which contains $N=747$ sequences of various length. A sliding sequence of stride equal to 1 is used to generate the set of $\kappa$-mers and, by keeping only the unique $\kappa$-mers, the sets present different cardinalities, ranging from $k_{min}=4002$ to $k_{max}=219972647$. These sets are then embedded with MinHash or SparseHash and the full matrix of all pairwise similarities is generated by computing distances on the measurement vectors. Due to the computational infeasibility of computing exact Jaccard values to estimate the quality of the approximation produced by MinHash and SparseHash, we approximate them using a large number of SparseHash measurements ($m=5\cdot 10^6$) and MinHash measurements ($m=10^6$) and then average the two estimates. This is done to avoid any bias for a specific algorithm. We remark that we use the MASH code provided by \cite{Ondov2016} to implement MinHash. Due to the large values of $k$ and $m$, that code uses the bottom-$m$ MinHash sketch to accelerate the computation of the sketches. However, it is not amenable to binarization: each MinHash measurement requires 64 bits against 1 bit required by SparseHash. Figure \ref{fig:metagenome} shows the mean square error (MSE) on the matrix with all Jaccard pairwise similarities with respect to the true Jaccard values as function of the computation time required to compute all pairwise distances in the embedded space. It can be noticed that SparseHash outperforms MASH thanks to the 1-bit measurements and the faster $\mathrm{sim}_{\sf{sh}}$ and Hamming distance metrics that can be implemented with bitwise operations. 

\section{Conclusion}\label{sec:conclusions}
In this letter, we have analyzed SparseHash, a novel embedding technique for dimensionality reduction of sets. Exploiting the concept of sparsity and concentration results, SparseHash is proven to preserve Jaccard metric. Efficient implementations and numerical experiments show that SparseHash outperforms MinHash in different applications. Future work will envisage the comparison with other strategies, e.g., Bloom filters.

\bibliographystyle{plain}

\appendix
\section{Appendix}
In this Appendix, we provide the details of the proofs of Theorem 1 and Proposition 1. Here, we report all the steps.

\subsection{Notation}
To facilitate the reading, we recall the notation used in the manuscript.
\begin{itemize}
    \item $u,v\in\R^n$;
    \item $A\in\R^{m,n}$ is a $\gamma$-sparsified random matrix, $\gamma\in(0,1)$, that is: each $A_{i,j}$ is zero with probability $1-\gamma$, while  $A_{i,j}$ is generated according to given distribution with zero mean and finite variance with probability $\gamma$;
    \item $\mathds{1}({\{A\}})$ is the indicator function, which returns 1 when $A$ is true;
    \item $\mathds{1}({\{A,B\}})$ is the indicator function which returns 1 when both A and B are true;
    \item $S_u =\{i\in\{1,\dots,n\} \text{ s.t. } u_i\neq 0\} $ 
    \item $y,z\in\{0,1\}^m$
    \item $\text{sim}_{\cup}(y,z):=\frac{1}{m}\sum_{i=1}^m\mathds{1}({\{y_i=0,z_i=0\}})$;
\item $\text{sim}_{\cap}(y,v):=\frac{\sum_{i=1}^m\mathds{1}({\{y_i=0\})\sum_{j=1}^m\mathds{1}(\{z_j=0\}})}{m\sum_{i=1}^m\mathds{1}({\{y_i=0,z_i=0\}})}$;
    \item $J_{u,v}$ (or $J$) $=$ Jaccard coefficient of the pair $(u,v)$.
\end{itemize}

\subsection{Proof of Theorem 1}
We start with some preliminary properties.
\begin{lemma}\label{lemma:cup}Let $u\in\Sigma_{k_1}, v\in\Sigma_{k_2}$,
\begin{equation}\label{eq:1}
\mathbb{P}\left(\left|\mathrm{sim}_{\cup}(Au,Av)-(1-\gamma)^{\frac{k_1+k_2}{1+J_{u,v}}}]\right|>\epsilon\right)\leq 2\e^{-2\epsilon^2m}.
\end{equation}
\end{lemma}
\begin{proof}

$S_u=\mathrm{supp}(u)$ and $S_v=\mathrm{supp}(S_v)$
It should be noticed that $\mathds{1}({\{y_i=0,z_i=0\}})$ is a Bernoulli random variable with
\begin{align*}
\mathbb{E}[\mathrm{sim}_{\cup}(Au,Av)]&=(1-\gamma)^{|S_u\cup S_v|}\\
&=(1-\gamma)^{(k_1+k_2)/(1+J_{u,v})},
\end{align*}
where the last equality is obtained using the relation $$J_{u,v}=\frac{|S_u|+|S_v| }{|S_u\cup S_v|}-1.$$
By Hoeffding's inequality (see \cite{citeulike:3392582}) we have
\begin{align*}
\mathbb{P}\left(\left|\mathrm{sim}_{\cup}(Au,Av)-\mathbb{E}[\mathrm{sim}_{\cup}(Au,Av)]\right|>\epsilon\right)\leq 2\e^{-2\epsilon^2m}.
\end{align*}

\end{proof}

\begin{lemma}\label{lemma:product}
Let $X$ and $Z$ be two random variables such that
\begin{align*}
\mathbb{P}(|X-\mu_X|>\epsilon)&\leq p_X(\epsilon)\\
\mathbb{P}(|Z-\mu_Z|>\epsilon)&\leq p_Z(\epsilon)
\end{align*}
with $\mu_X>0, \mu_Z>0$, then
\begin{align*}
\mathbb{P}(|XZ-\mu_X\mu_Z|>\epsilon)&\leq p_X\left(\frac{\epsilon}{3\mu_Z}\right)\\&+p_Z\left(\min\left\{\frac{\epsilon}{2\mu_X},\frac{\mu_Z}{2}\right\}\right)
\end{align*}
\end{lemma}
\begin{proof}Let us define the following events
\begin{align*}
    E_X&=\left\{|X-\mu_X|\leq\frac{\epsilon}{3\mu_Z}\right\}\\
    E_Z&=\left\{|Z-\mu_Z|\leq\min\left\{\frac{\epsilon}{2\mu_X},\frac{\mu_Z}{2}\right\}\right\}.
\end{align*}
From the law of total probability the following series of inequalities follows:
\begin{align*}
\mathbb{P}(|XZ-\mu_X\mu_Z|>\epsilon)&=\mathbb{P}(|XZ-\mu_X\mu_Z|>\epsilon|E_X)\mathbb{P}(E_X)\\
&+\mathbb{P}(|XZ-\mu_X\mu_Z|>\epsilon|E_X^{\mathrm{c}})\mathbb{P}(E_X^{\mathrm{c}})\\
&\leq\mathbb{P}(|XZ-\mu_X\mu_Z|>\epsilon|E_X)+\mathbb{P}(E_X^{\mathrm{c}})
\end{align*}
Repeating the same argument, we get
\begin{align*}
&\mathbb{P}(|XZ-\mu_X\mu_Z|>\epsilon)\leq\\
&~\leq\mathbb{P}(|XZ-\mu_X\mu_Z|>\epsilon|\color{black}{E_X\cap E_Z})+\mathbb{P}(E_X^{\mathrm{c}})+\mathbb{P}(E_Z^{\mathrm{c}})
\end{align*}
It should be noticed that
\begin{align*}
|XZ-\mu_X\mu_Z|&=|XZ-\mu_X Z+\mu_X Z-\mu_X\mu_Z|\\
&\leq|Z|~|X-\mu_X|+\mu_X |Z-\mu_Z|.
\end{align*}
If $E_Z, E_X$ hold then 
$   \mu_Z/2\leq Z\leq 3\mu_Z/2 $ and 
$$|XZ-\mu_X\mu_Z|\leq \epsilon
$$
from which $$\mathbb{P}(|XZ-\mu_X\mu_Z|>\epsilon|\color{black}{E_X\cap E_Z})=0.$$
We conclude that
\begin{align*}
&\mathbb{P}(|XZ-\mu_X\mu_Z|>\epsilon)\\
&~\leq\mathbb{P}(E_X^{\mathrm{c}})+\mathbb{P}(E_Z^{\mathrm{c}})\\
&~=p_X\left(\frac{\epsilon}{3\mu_Z}\right)+p_Z\left(\min\left\{\frac{\epsilon}{2\mu_X},\frac{\mu_Z}{2}\right\}\right).
\end{align*}
\end{proof}
From Lemma \ref{lemma:product}, we can derive the following result.
\begin{corollary}\label{corol:1}Let $u\in\Sigma_{k_1}, v\in\Sigma_{k_2}$, and
denote 
\begin{align*}
X&=\frac{\sum_{i=1}^m\mathds{1}(\{(Au)_i=0\})}{m}\\
Z&=\frac{\sum_{i=1}^m\mathds{1}(\{(Av)_i=0\})}{m}
\end{align*}
then
\begin{equation}\label{eq:2}
\mathbb{P}\left(\left|XZ-(1-\gamma)^{k_1+k_2}\right|>\epsilon\right)\leq 4\e^{-2m\min\left\{\frac{\epsilon^2}{9},\frac{(1-\gamma)^{2k_{\max}}}{4}\right\}}
\end{equation}\end{corollary} 

\begin{proof}
It should be noticed that $\mathds{1}(\{(Au)_i=0\})$ is a Bernoulli random variable with $\mathbb{P}(\{(Au)_i=0\})=(1-\gamma)^{k_1}:=\mu_X$. Then from the Hoeffding's inequality \cite{citeulike:3392582} we get
$$\mathbb{P}\left(\left|X-\mu_X\right|>\epsilon\right)\leq 2\e^{-2\epsilon^2m}.
$$
and, analogously denoting $\mu_Z=(1-\gamma)^{k_2}$,
$$\mathbb{P}\left(\left|Z-\mu_Z\right|>\epsilon\right)\leq 2\e^{-2\epsilon^2m}.
$$
Applying Lemma \ref{lemma:product} and computing
\begin{equation}\label{eq:3}
\mathbb{P}\left(\left|X-\mu_X\right|>\frac{\epsilon}{3\mu_Z}\right)\leq 2\e^{-\frac{2\epsilon^2m}{9(1-\gamma)^{2k_2}}}\leq 2\e^{-\frac{2\epsilon^2m}{9}}
\end{equation} 
and, similarly,
\begin{equation}
\begin{split}\label{eq:4}
&\mathbb{P}\left(\left|Z-\mu_Z\right|>\min\left\{\frac{\epsilon}{2\mu_X},\frac{\mu_Z}{2}\right\}\right)\\
&~\leq 2\e^{-2m\min\left\{\frac{\epsilon^2}{4(1-\gamma)^{\color{black}k_1}},\frac{(1-\gamma)^{2k_2}}{4}\right\}}\\
&~\leq 2\e^{-2m\min\left\{\frac{\epsilon^2}{4},\frac{(1-\gamma)^{2k_{\max}}}{4}\right\}}.\end{split}
\end{equation} 
we obtain \eqref{eq:2}.
\end{proof}

\begin{lemma}\label{lemma:ratio}
Let $X,Z$ be random variables such that
\begin{align*}
\mathbb{P}\left(|X-\mu_X|>\epsilon\right)\leq p_X(\epsilon), ~
\mathbb{P}\left(|Z-\mu_Z|>\epsilon\right)&\leq p_Z(\epsilon)
\end{align*}
with $\mu_X>0, \mu_Z>0$, then
\begin{equation}\begin{split}
&\mathbb{P}\left(\left|\frac{X}{Z}-\frac{\mu_X}{\mu_Z}\right|>\epsilon\right)\\
&~\leq p_X\left(\frac{\epsilon \mu_Z}{4}\right)+p_Z\left(\min\left\{\frac{\epsilon}{4}\frac{\mu_Z^2}{\mu_X},\frac{\mu_Z}{2}\right\}\right),
\end{split}\end{equation}
\end{lemma}

\begin{proof}Let us define the following events
\begin{align*}
    E_X&=\left\{|X-\mu_X|\leq\frac{\epsilon\mu_Z}{4}\right\}\\
    E_Z&=\left\{|Z-\mu_Z|\leq\min\left\{\frac{\epsilon}{4}\frac{\mu_Z^2}{\mu_X},\frac{\mu_Z}{2}\right\}\right\}.
\end{align*}
From the law of total probability the following series of inequalities follows
\begin{align*}
\mathbb{P}\left(\left|\frac{X}{Z}-\frac{\mu_X}{\mu_Z}\right|>\epsilon\right)&=
\mathbb{P}\left(\frac{|X\mu_Z-Z\mu_X|}{|Z||\mu_Z|}>\epsilon\right)\\
&=
\mathbb{P}\left(\frac{|X\mu_Z-\mu_X\mu_Z+\mu_X\mu_Z-Z\mu_X|}{|Z|\mu_Z}>\epsilon\right)\\
&\leq
\mathbb{P}\left(\frac{\mu_Z|X-\mu_X|+\mu_X|Z-\mu_Z|}{|Z|\mu_Z}>\epsilon\right)\\
&\leq
\mathbb{P}\left(\frac{\mu_Z|X-\mu_X|+\mu_X|Z-\mu_Z|}{|Z|\mu_Z}>\epsilon\bigg|  \color{black}{E_X\cap E_Z} \right)\\
& +\mathbb{P}\left(E_X^{\mathrm{c}}\right)+\mathbb{P}\left(E_Z^{\mathrm{c}}\right)
\end{align*}
If $E_Z$ holds then $Z\geq\mu_Z/2$ and
$$
\mathbb{P}\left(\frac{\mu_Z|X-\mu_X|+\mu_X|Z-\mu_Z|}{|Z||\mu_Z|}>\epsilon\bigg|  E_X,E_Z \right)=0,
$$
we have
\begin{align*}
&\mathbb{P}\left(\left|\frac{X}{Z}-\frac{\mu_X}{\mu_Z}\right|>\epsilon\right) \leq p_X\left(\frac{\epsilon\mu_Z}{4}\right)+p_Z\left(\min\left\{\frac{\epsilon}{4}\frac{\mu_Z^2}{\mu_X},\frac{\mu_Z}{2}\right\}\right).
\end{align*}
\end{proof}
From Lemma \ref{lemma:ratio} and Corollary \ref{corol:1}, we obtain the following result.
\begin{corollary}\label{corol:2}
Let $u\in\Sigma_{k_1}, v\in\Sigma_{k_2}$, then
\begin{align}\begin{split}\label{eq:5}
&\mathbb{P}\left(\left|\mathrm{sim}_{\cap}(Au,Av)-(1-\gamma)^{\frac{(k_1+k_2)J_{u,v}}{1+J_{u,v}}}]\right|>\epsilon\right)\\
&\leq 6\exp\left({-m\min\left\{(1-\gamma)^{8k_{\max}}\frac{\epsilon^2}{8},\frac{(1-\gamma)^{4k}}{2}\right\}}\right)
\end{split}
\end{align}
\end{corollary}

\begin{proof}The inequality in \eqref{eq:5} follows directly from Lemma \ref{lemma:ratio} by setting 
\begin{gather*}
X=\frac{\sum_{i=1}^m\mathds{1}(\{(Au)_i=0\})\sum_{i=1}^m\mathds{1}(\{(Av)_i=0\})}{m^2},
\\ 
Z=\mathrm{sim}_{\cup}(Au,Av),\\
\mu_X=(1-\gamma)^{k_1+k_2},~
\mu_Z=(1+\gamma)^{(k_1+k_2)/(1+J_{u,v})}
\end{gather*}
and estimating the probabilities using Lemma \ref{lemma:ratio}, Corollary \ref{corol:1}, $k_1+k_2\leq 2k_{\max}$ and $J\geq0$
$$
p_X\left(\frac{\epsilon \mu_Z}{4}\right)\leq 4\e^{-m\min\left\{\frac{\epsilon^2}{9\cdot 8}(1-\gamma)^{4k_{\max}},\frac{(1-\gamma)^{2k_{\max}}}{2}\right\}}
$$
and
\begin{align*}
    &p_Z\left(\min\left\{\frac{\epsilon}{4}\frac{\mu_Z^2}{\mu_X},\frac{\mu_Z}{2}\right\}\right)=\\
    &= 2\exp\left({-2m\left(\min\left\{\frac{\epsilon(1-\gamma)^{2(k_1+k_2)/(1+J)}}{4(1-\gamma)^{k_1+k_2}},\frac{(1-\gamma)^{(k_1+k_2)/(1+J)}}{2}\right\}\right)^2}\right)\\
    &\leq 2\exp\left({-2m\left(\min\left\{(1-\gamma)^{4k_{\max}}\frac{\epsilon}{4},\frac{(1-\gamma)^{2k_{\max}}}{2}\right\}\right)^2}\right)\\
    &= 2\exp\left({-m\min\left\{(1-\gamma)^{8k_{\max}}\frac{\epsilon^2}{8},\frac{(1-\gamma)^{4k_{\max}}}{2}\right\}}\right).
\end{align*}
We conclude that
\begin{align*}
&\mathbb{P}\left(\left|\mathrm{sim}_{\cap}(Au,Av)-(1-\gamma)^{\frac{(k_1+k_2)J_{u,v}}{1+J_{u,v}}}\right|>\epsilon\right)\\
&~\leq p_X\left(\frac{\epsilon \mu_Z}{4}\right)+p_Z\left(\min\left\{\frac{\epsilon}{4}\frac{\mu_Z^2}{\mu_X},\frac{\mu_Z}{2}\right\}\right)\\
&~\leq 6\exp\left({-m\min\left\{(1-\gamma)^{8k_{\max}}\frac{\epsilon^2}{8},\frac{(1-\gamma)^{4k_{\max}}}{2}\right\}}\right).
\end{align*}
\end{proof}

\begin{lemma}\label{lemma:log}
Let $X$ be a positive random variable such that
\begin{align*}
\mathbb{P}\left(|X-\mu_X|>\epsilon\right)\leq p_X(\epsilon), ~
\end{align*}
with $\mu_X>0$, then
$$
\mathbb{P}\left(\left|\log(X)-\log(\mu_X)\right|>\epsilon\right)\leq p_X\left(\epsilon\mu_X\right).
$$
\end{lemma}

\begin{proof}
{\color{black}Noticing that $|\log(1+|t|)|\leq |t|$ for any $t$, the following series of relations holds
\begin{align*}
&\mathbb{P}\left(\left|\log(X)-\log(\mu_X)\right|>\epsilon\right)\leq\mathbb{P}\left(\left|\log\left(1+\left|\frac{X-\mu_X}{\mu_X}\right|\right)\right|>\epsilon\right)
\\
&~\leq\mathbb{P}\left(\frac{|X-\mu_X|}{\mu_X}>\epsilon\right)=p_X(\epsilon\mu_X).
\end{align*}}
\end{proof}

\subsubsection*{Conclusion of the proof of Theorem 1}

Applying Lemma \ref{lemma:log} to the results obtained in Lemma \ref{lemma:cup} and \ref{corol:2} we obtain:
\begin{align*}
&\mathbb{P}\left(\left|\log(\mathrm{sim}_{\cap}(Au,Av))-\frac{J_{u,v}(k_1+k_2)\log(1-\gamma)}{1+J_{u,v}}\right|>\epsilon\right)\\
&\leq 6\exp\left({-m\min\left\{(1-\gamma)^{8k_{\max}}\frac{\epsilon^2\mu_Z^2}{8},\frac{(1-\gamma)^{4k_{\max}}}{2}\right\}}\right)\\
& =6\exp\left({-m\min\left\{(1-\gamma)^{8k_{\max}}\frac{\epsilon^2(1-\gamma)^{2(k_1+k_2)J/(1+J)}}{8},\frac{(1-\gamma)^{4k_{\max}}}{2}\right\}}\right)\\
& \leq 6\exp\left({-m\min\left\{(1-\gamma)^{12k_{\max}}\frac{\epsilon^2}{8},\frac{(1-\gamma)^{4k_{\max}}}{2}\right\}}\right)\\
&= 6\exp\left({-m(1-\gamma)^{12k_{\max}}\frac{\epsilon^2}{8}}\right);\\
&\mathbb{P}\left(\left|\log(\mathrm{sim}_{\cup}(Au,Av))-\frac{(k_1+k_2)\log(1-\gamma)}{1+J_{u,v}}\right|>\epsilon\right)\\
&\leq 2\exp\left({-2m\epsilon^2(1-\gamma)^{2(k_1+k_2)}}\right)\leq 2\exp\left({-2m\epsilon^2(1-\gamma)^{4k_{\max}}}\right).
\end{align*}
Finally, using Lemma \ref{lemma:ratio} and the assumption $
m> 32\frac{\log 4+\beta\log N}{\gamma^2k^2_{\min}\e^{-\gamma k_{\max}}\epsilon^2}
$,
we get
\begin{align*}
&\mathbb{P}\left(\left|\frac{\log(\mathrm{sim}_{\cap}(Au,Av))}{\log(\mathrm{sim}_{\cup}(Au,Av))}-J_{u,v}\right|>\epsilon\right)\leq\\
&~\leq 8\exp\left({-m\frac{\epsilon^2}{128}(1-\gamma)^{12 k_{\max}}4 k^2_{\min}\log^{2}(1-\gamma)}\right)\leq N^{-\beta+2}.
\end{align*}
The thesis follows by applying the union bound on the last expression.
\subsection{Proof of Proposition 1}

If a pair of signals in $\mathcal{X}$ has Jaccard coefficient $J\in[0,1]$, then the probability that their hashes computed with SparseHash become a candidate pair is given by:
$$P_{\text{sh}}=1-(1-p_{\text{sh}}^{m_{\text{sh}}})^{L_{\text{sh}}}$$
with $p_{\text{sh}}=2^{\frac{J-1}{1+J}}$, while 
for 1-bit MinHash,  
$$P_{\text{mh}}=1-(1-p_{\text{mh}}^{m_{\text{mh}}})^{L_{\text{mh}}}$$ 
with $p_{\text{mh}}=(J+1)/2$.

Let $m_{sh}=m_{mh}=m$. From the previous equations, it is straightforward to prove that: $$\frac{L_{\text{sh}}}{L_{\text{mh}}}
  =\frac{\log\left(1-\left((J+1)/2\right)^{m}\right)}{\log\left(1-2^{m\frac{J-1}{1+J}}\right)}.$$
  
  We then obtain $L_{sh}\leq L_{mh}$ if $$\frac{\log\left(1-\left((J+1)/2\right)^{m}\right)}{\log\left(1-2^{m\frac{J-1}{1+J}}\right)}\leq 1.$$

\begin{figure}[h]
    \centering
    \includegraphics[width=0.7\columnwidth]{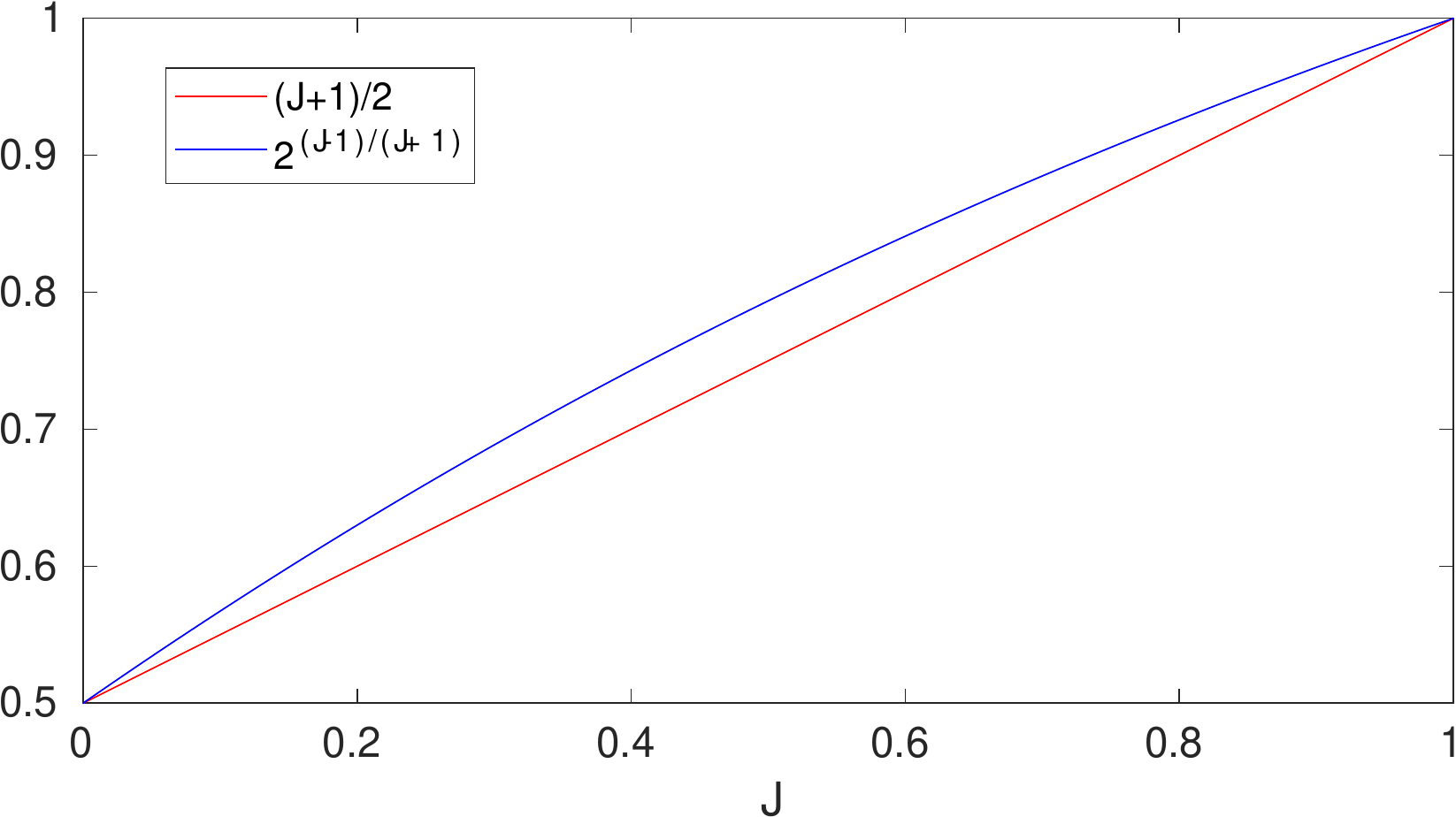}
    \caption{Proof of Proposition 1}
    \label{fig1}
\end{figure}

Since $\log\left(1-2^{m\frac{J-1}{1+J}}\right)<0$,
\begin{equation}
\begin{split}
    &\log\left(1-\left((J+1)/2\right)^{m}\right)\geq\log\left(1-2^{m\frac{J-1}{1+J}}\right)\\
    &\left(1-\left((J+1)/2\right)^{m}\right)\geq\left(1-2^{m\frac{J-1}{1+J}}\right)\\
    &1-(J+1)/2\geq 1-2^{\frac{J-1}{1+J}}\\
    &(J+1)/2\leq 2^{\frac{J-1}{1+J}}.
\end{split}    
\end{equation}
We can then numerically verify that the last inequality is always true (see Figure \ref{fig1}), in particular the strict inequality holds for any $J\in(0,1)$.

\end{document}